\documentclass[lettersize,journal]{IEEEtran}
\usepackage{amsmath,amsfonts}
\usepackage{gensymb}

\usepackage{algorithmic}
\usepackage{algorithm}
\usepackage{array}
\usepackage[caption=false]{subfig}
\usepackage{booktabs}
\usepackage{textcomp}
\usepackage{stfloats}
\usepackage{url}
\usepackage{verbatim}
\usepackage{graphicx}
\usepackage{cite}
\usepackage{multirow} 
\begin{document}

\title{Spatial Analysis and Synthesis Methods:\\ Subjective and Objective Evaluations Using Various Microphone Arrays in the Auralization of a Critical Listening Room}

\author{Alan Pawlak,~Hyunkook Lee,~Aki Mäkivirta,~and Thomas Lund 
\thanks{The research presented in this paper was funded by Genelec Oy and the University of Huddersfield.}
\thanks{A. Pawlak and H. Lee are with the Applied Psychoacoustics Lab (APL), University of Huddersfield,
HD1 3DH Huddersfield, United Kingdom (e-mail: alan.pawlak@hud.ac.uk; h.lee@hud.ac.uk)}
\thanks{A. Mäkivirta and T. Lund are with Genelec OY, 74100 Iisalmi, Finland (e-mail: aki.makivirta@genelec.com; thomas.lund@genelec.com)}
}

\markboth{In peer review for publication in the IEEE/ACM Transactions on Audio, Speech, and Language Processing, submitted December 2023}%
{Shell \MakeLowercase{\textit{et al.}}: A Sample Article Using IEEEtran.cls for IEEE Journals}


\maketitle
\begin{abstract}
  Parametric sound field synthesis methods, such as the Spatial Decomposition Method (SDM) and Higher-Order Spatial Impulse Response Rendering (HO-SIRR), are widely used for the analysis and auralization of sound fields. This paper studies the performances of various sound field synthesis methods in the context of the auralization of a critical listening room. The influence on the perceived spatial and timbral fidelity of the following factors is considered: the rendering framework, direction of arrival (DOA) estimation method, microphone array structure, and use of a dedicated center reference microphone with SDM. Listening tests compare the synthesized sound fields to a reference binaural rendering condition. Several acoustic parameters are measured to gain insights into objective differences between methods. A high-quality pressure microphone improves the SDM framework's timbral fidelity. Additionally, SDM and HO-SIRR show similarities in spatial fidelity. Performance variation between SDM configurations is influenced by the DOA estimation method and microphone array construction. The binaural SDM (BSDM) presentations display temporal artifacts impacting sound quality.
\end{abstract}

\section{Introduction}

\IEEEPARstart{D}{irectional} sound field analysis reveals spatial reflection patterns in enclosed spaces. Its goals include: (i) pinpointing problematic room reflections \cite{abdouSpatialInformationSound1996}, (ii) understanding perceptual impacts of room features \cite{abdouSpatialInformationSound1996}, (iii) identifying reflections contributing to the desirable perceptual attributes \cite{goverMeasurementsDirectionalProperties2004}, (iv) assessing the impact of reflection regions on objective metrics \cite{yamasakiMeasurementSpatialInformation1989}, (v) advancing psychoacoustic research \cite{goverMeasurementsDirectionalProperties2004}, and (vi) forming the foundation for parametric auralisation systems \cite{merimaaSpatialImpulseResponse2005,tervoSpatialDecompositionMethod2013}.

Over the past two decades, the principles of spatial analysis have driven significant growth in parametric spatial audio rendering. This trend began with the introduction of the Spatial Impulse Response Rendering (SIRR) \cite{merimaaSpatialImpulseResponse2005}, which utilizes first-order spherical harmonics (SPH) for sound field analysis and synthesis in the time-frequency domain. A subsequent, more straightforward approach known as the Spatial Decomposition Method (SDM) was introduced by Tervo et al. \cite{tervoSpatialDecompositionMethod2013}. This method operates in the time domain and interprets each sample in an impulse response as an image source characterized by both pressure and direction. The public availability of SDM as a MATLAB toolbox \cite{tervoSDMToolbox2023} has made it a popular choice for analyzing enclosed spaces, auralization, and research \cite{patynenAnalysisConcertHall2013, tervoSpatialAnalysisSynthesis2015, gariPhysicalPerceptualComparison2016, amengualgariRealtimeAuralizationRoom2016, mullerPerceptualDifferencesModifications2022, bedernaPerceptualDetectionThresholds2023, gomesPerceptualConsequencesDirection2022}. As object-based audio gained prominence, the Reverberant Spatial Audio Object (RSAO) was developed, parameterizing spatial room impulse response (SRIR) into a concise set of coefficients. This was aimed at enabling reverberation synthesis within audio object renderers \cite{colemanObjectbasedReverberationSpatial2017}.

Later, the HO-SIRR, a higher-order adaptation of the SIRR, was introduced \cite{mccormackHigherOrderSpatialImpulse2020}, offering enhanced spatial resolution through the use of higher-order SPH. Concurrently, the Ambisonic SDM (ASDM) was introduced, allowing for the upscaling of the first-order ambisonics (FOA) to higher-order (HOA). Subsequent innovations include the binaural versions of SDM (BSDM) \cite{amengualgariOptimizationsSpatialDecomposition2021} and HO-SIRR\cite{holdParametricBinauralReproduction2022}, an enhanced ASDM known as 4D-ASDM \cite{hoffbauerFourDirectionalAmbisonicSpatial2022}, and the Reproduction and Parameterisation of Array Impulse Responses (REPAIR) \cite{mccormackSpatialReconstructionBasedRendering2023}.

Despite the rapid developments, the SDM and SIRR (and HO-SIRR) remain predominant. Previous studies, such as \cite{pulkkiSpatialImpulseResponse2006, tervoSpatialDecompositionMethod2013, mccormackHigherOrderSpatialImpulse2020,mccormackSpatialReconstructionBasedRendering2023}, have evaluated these methods primarily through subjective experiments with loudspeakers employing simulated environments. However, some acoustic simulation methods, like the Image Source Method (ISM) \cite{allenImageMethodEfficiently1979} may favor methods like the SDM due to shared underlying assumptions. Moreover, inconsistencies arise, such as those between the SDM and HO-SIRR evaluations, which may stem from different microphone array configurations or direction-of-arrival (DOA) estimation techniques used. Previous studies have not comprehensively compared all these methods under real-life conditions, which this study intends to do.

Additionally, the experimental setups in prior studies appear to be somewhat limited. While SDM has been used with various microphone arrays, only two studies have examined how the array affects SDM's auralization quality \cite{ahrensPerceptualEvaluationBinaural2019} and DOA accuracy \cite{amengualgariOptimizationsSpatialDecomposition2021}. Despite commonly using signals from any omnidirectional microphone in arrays, no perceptual differences were found between signals from Ambisonics and a central omni microphone, although the research lacked methodological detail \cite{amengualgariSpatialAnalysisAuralization2017}.

In this paper\footnote{This paper extends our initial study \cite{pawlakSubjectiveEvaluationSpatial2021}, offering a detailed analysis with more subjects and source positions, objective metrics, in-lab experiments (replacing the previous remote setup due to COVID-19), refined Eigenmike em32 impulse response measurements, and revising evaluated systems.}, we present a comprehensive perceptual evaluation to test the hypothesis that different synthetic binaural room impulse responses (BRIRs) produced using various parametric spatial audio rendering techniques---specifically SDM, BSDM, and HO-SIRR---will differ in spatial and timbral fidelities compared to a reference BRIR recorded with the KU100 dummy head. Additionally, building on \cite{tervoSpatialDecompositionMethod2013} and \cite{ahrensPerceptualEvaluationBinaural2019}, we investigate whether a greater number of sensors in an array leads to improved localization performance. To this end, we employ both an Eigenmike em32---equipped with 32 omni capsules on a rigid sphere with a 42 mm radius---and a compact microphone array with six omni microphones for SDM. Further, we examine the impact of direction of arrival (DOA) estimation algorithms in SDM, focusing on those based on time difference of arrival (TDOA) and pseudo intensity vectors (PIVs), and assess the influence of using a dedicated omnidirectional microphone at the array's center as a pressure signal in SDM.

The study is confined to a room that complies with recommendations in ITU-R 1116-3 \cite{itu-rbs.1116-3MethodsSubjectiveAssessment2015} since our main focus is on the spatial analysis and auralization of critical listening or sound mixing room. We operate under the assumption that our findings will be replicable in similar rooms adhering to this widely recognized standard. Our study utilizes source positions from six orientations, aligned with industry standards like the ITU-R BS.2051-2 \cite{itu-rbs.2051-3AdvancedSoundSystem2022} and Dolby Atmos 7.1.4 configurations. The aim is to ground the discussion in scientific rigor amidst the rapid expansion of parametric spatial audio reproduction techniques, delineating current standings, identifying gaps, and defining a system for optimal spatial data capture and accurate reproduction in the context of critical listening room.

The paper is structured as follows: Section \ref{sec:doa} introduces the foundational theories of the direction of arrival estimation, with a focus on time difference of arrival (TDOA) and Pseudo Intensity Vectors (PIVs) which are crucial to the SDM and SIRR methods. Section \ref{sec:systems} elaborates on the specifics of the SDM and SIRR methods. Section \ref{sec:method} details our methodology and experimental design. Section \ref{sec:subjective} showcases our perceptual study results, based on the MUSHRA methodology. Section \ref{sec:objective} reports on objective metrics. Section \ref{sec:discussion} synthesizes our key findings. Section \ref{sec:conclusion} concludes the paper.

\section{Direction of Arrival Estimation Methods} \label{sec:doa}
Estimating the direction of arrival (DOA) is a fundamental aspect of parametric sound field synthesis methods, as it determines the wave's origin and propagation direction. Various methods have been developed for DOA estimation, including Time Difference of Arrival (TDOA)-based methods like the Generalized Cross Correlation (GCC), subspace methods such as ESPRIT or MUSIC, beamforming approaches, and Pseudo Intensity Vectors (PIVs), with comprehensive tutorials available in \cite{jarrettTheoryApplicationsSpherical2017}. This section will focus on two approaches particularly relevant to SDM and SIRR: TDOA and PIVs.

\subsection{Time Difference of Arrival (TDOA)} \label{sec:tdoa}
The time difference of arrival is frequently used to determine the source's direction of arrival. It measures the time lag of a signal across multiple sensors. Knowing the sensors' relative positions and these time lags allows estimation of the source's origin direction. Common TDOA estimation techniques include cross-correlation with weightings such as GCC-SCOT and GCC-PHAT \cite{knappGeneralizedCorrelationMethod1976}.

While the Spatial Decomposition Method (SDM) can be combined with any appropriate direction of arrival (DOA) estimation algorithm, the foundational paper utilized a least squares method for DOA estimation via TDOA using GCC with no weighting \cite{tervoSpatialDecompositionMethod2013}. As this algorithm was incorporated into a popular toolbox \cite{tervoSDMToolbox2023}, it is often linked with the original SDM. For detailed equations and further technical specifics, please refer to \cite{tervoSpatialDecompositionMethod2013, zhangCrossCorrelationBaseddiscrete2005, knappGeneralizedCorrelationMethod1976, yli-hietanenLowcomplexityAngleArrival1996}.

\subsection{Pseudo-Intensity-Vectors (PIVs)} \label{sec:piv}

Pseudo Intensity Vectors (PIVs) offer a viable alternative to TDOA-based DOA estimation, building upon historical sound intensity measurement methodologies \cite{yamasakiApplicationDigitalTechniques1978, sekiguchiAnalysisSoundField1992, abdouSpatialInformationSound1996, tervoEstimationReflectionsImpulse2011}. The term "Pseudo Intensity Vectors" was first introduced by Jarrett et. al. \cite{jarrett3DSourceLocalization2010} and refers to the sound intensity vectors computed from the zeroth and first-order eigenbeams.

The subjective results obtained by McCormack et al.\cite{mccormackHigherOrderSpatialImpulse2020} and Ahrens \cite{ahrensAuralizationOmnidirectionalRoom2019} implied that broadband DOA estimation using PIVs is inferior to other SDM and SIRR configurations in auralization. However, Zaunschirm et al. showed that using this method for bandwidth between 200 Hz and just below the microphone array's spatial aliasing frequency resulted in SDM rendering nearly identical to the binaural reference in terms of auditory image width, distance, and diffuseness \cite{zaunschirmBRIRSynthesisUsing2018}. Bassuet stressed minimizing microphone directivity effects in broadband PIV-based DOA estimation, suggesting a 100 Hz–5 kHz filter range where the Soundfield first-order Ambisonics microphone exhibits consistent directivity characteristics \cite{bassuetNewAcousticalParameters2011}.

Alternatively, the PIV can be represented in the frequency domain, as done in SIRR and HO-SIRR \cite{merimaaSpatialImpulseResponse2005, mccormackHigherOrderSpatialImpulse2020}, the PIV is computed for each frequency bin, averaged over time windows. This method allows for the calculation of the diffuseness coefficient $\psi(\omega)$, indicating the balance between the pseudo-intensity vector magnitude and the energy density (details in \cite{merimaaSpatialImpulseResponse2005, mccormackHigherOrderSpatialImpulse2020}).

\section{Parametric Sound Reproduction Methods} \label{sec:systems}

\subsection{Spatial Decomposition Method (SDM)}
The Spatial Decomposition Method (SDM), introduced in 2013, utilizes the image source model for parametric spatial encoding, treating impulse response samples as broadband image sources \cite{tervoSpatialDecompositionMethod2013}. The resulting metadata can be utilized for loudspeaker reproduction or binaural reproduction using Head-Related Transfer Functions (HRTFs). SDM's process involves a microphone array for direction of arrival (DOA) estimation and an omnidirectional microphone for pressure signal capture. The method comprises two stages: spatial analysis using DOA estimation algorithms (TDOA or PIVs) and synthesis, which utilizes DOA data and pressure signals to create directional output signals using techniques like Vector Base Amplitude Panning (VBAP) \cite{tervoSpatialDecompositionMethod2013, patynenAmplitudePanningDecreases2014, amengualgariRealtimeAuralizationRoom2016}, K-Nearest Neighbour (KNN) mapping \cite{patynenAmplitudePanningDecreases2014, tervoSpatialAnalysisSynthesis2015, ahrensAuralizationOmnidirectionalRoom2019}, or Ambisonics \cite{zaunschirmBinauralRenderingMeasured2020}. Post-equalization to mitigate spectral whitening and optimized loudspeaker grids are also utilized for enhanced sound reproduction \cite{tervoSpatialAnalysisSynthesis2015, puomioOptimizationVirtualLoudspeakers2017}.

Binaural Spatial Decomposition Method (BSDM), offers improvements for binaural reproduction \cite{amengualgariOptimizationsSpatialDecomposition2021}. It includes the rotation of the DOA matrix for various head orientations and post-processing techniques for direct sound enforcement and DOA quantization. BSDM also features RTMod+AP equalization for better echo density and decay, building on the previous post-equalization method designed for loudspeaker reproduction \cite{tervoSpatialAnalysisSynthesis2015}. Later, updates to the publicly available toolbox have introduced features such as impulse response denoising and band-limited spatial analysis \cite{amengualgariBinauralSDM2023}.

SDM's performance hinges on the microphone array configuration and window size. Historically, SDM has employed a variety of microphone arrays, from GRAS 50VI probes to custom arrays with varying spacings \cite{tervoSpatialDecompositionMethod2013,patynenAnalysisConcertHall2013,tervoPreferencesCriticalListening2014,tervoSpatialAnalysisSynthesis2015, ahrensPerceptualEvaluationBinaural2019, amengualgariSpatialAnalysisAuralization2017,gariPhysicalPerceptualComparison2016,gariFlexibleBinauralResynthesis2019,amengualgariOptimizationsSpatialDecomposition2021}. Optimal array design, guided by research, focuses on specific microphone spacings to minimize DOA errors and perceptual discrepancies \cite{tervoSpatialDecompositionMethod2013, ahrensPerceptualEvaluationBinaural2019, amengualgariOptimizationsSpatialDecomposition2021}. PIV-based DOA estimation allows flexibility in microphone array choice, but it requires an array capable of producing high-quality first-order spherical harmonics \cite{amengualgariOptimizationsSpatialDecomposition2021}. In this method, windowing aims at smoothing the DOA while in TDOA-based estimation, the window size is key to performance, ensuring a balance between temporal and spatial resolution. Larger windows enhance estimation robustness but increase the risk of multiple reflections, conflicting with the single-reflection assumption \cite{tervoSpatialDecompositionMethod2013}. Historically, window sizes were chosen arbitrarily \cite{tervoSpatialAnalysisSynthesis2015, tervoPreferencesCriticalListening2014,patynenAnalysisConcertHall2013, patynenAmplitudePanningDecreases2014}. Recent research by Amengual Garí et al. on optimal window size for DOA estimation in a simulated setup suggests that for a 100 mm spaced array, 36 or 64 samples at 48 kHz is most effective \cite{amengualgariOptimizationsSpatialDecomposition2021}.

\subsection{Spatial Impulse Response Rendering (SIRR)}
Spatial Impulse Response Rendering (SIRR) is the first parametric spatial encoding method for SRIR, focusing on emulating key perceptual features like Interaural Time Difference (ITD), Interaural Level Difference (ILD), and Interaural Coherence (IC) \cite{merimaaSpatialImpulseResponse2005}. Distinct from the SDM, SIRR operates in the time-frequency domain using spherical harmonic input. 

The processing in SIRR utilizes the Short Time Fourier Transform (STFT) with a Hann window, using Pseudo Intensity Vectors (PIVs)---as outlined in Section \ref{sec:piv}---to determine the direction of arrival (DOA) and diffuseness for each time-frequency bin. In the synthesis stage, the pressure response is divided into non-diffuse and diffuse parts, guided by the estimated diffuseness coefficient. The directional part is panned using Vector-Based Amplitude Panning (VBAP), while the diffuse component is decorrelated and distributed uniformly among the loudspeakers.

SIRR was extended to Higher-Order Spatial Impulse Response Rendering (HO-SIRR) \cite{mccormackHigherOrderSpatialImpulse2020}, which employs higher-order spherical harmonics and beamforming to divide the sound field into uniform sectors. Each sector undergoes separate analysis, allowing for a more precise estimation. This is especially valuable in challenging situations, for instance, when sound events from two distinct directions reach the array simultaneously. During the synthesis stage, the directional components from all sectors are panned via VBAP to their respective channels and then combined. Meanwhile, the diffuse components are re-encoded into spherical harmonics, decoded to loudspeaker signals, and then decorrelated.

The binaural variant of HO-SIRR, as detailed by Hold et al. \cite{holdParametricBinauralReproduction2022}, addresses coloration and HRTF resolution in virtual speaker binauralization. This approach integrates HRTFs in the synthesis phase of HO-SIRR, rendering directional components according to arrival directions and diffuse components by their sector steering directions. Objective analyses have demonstrated a reduction in coloration compared to the traditional HO-SIRR method.

\section{Experimental Design} \label{sec:method}

\subsection{Evaluation Method}
The main goal of this study is to determine which synthetic binaural room impulse response (BRIR) yields an auralization most perceptually similar to that produced by a BRIR recorded with the KU100 dummy head. To evaluate this similarity, we adopted the fidelity attribute, as defined by \cite{zielinskiComparisonBasicAudio2005}. Zieliński described it as the "trueness of reproduction quality to that of the original". The experiment had two dependent variables: (i) spatial fidelity and (ii) timbral fidelity.

The experiment used the MUSHRA methodology \cite{itu-rbs.1534-3MethodSubjectiveAssessment2014}, in which participants rated the similarity of test sounds to a reference on a continuous scale from 1.0 ("Extremely different") to 5.0 ("Same"), with intermediate values of 2.0 ("Very different"), 3.0 ("Different"), and 4.0 ("Slightly different"). This scale has been used in analogous subjective studies employing the MUSHRA methodology \cite{hiyamaMinimumNumberLoudspeakers2002, tervoSpatialDecompositionMethod2013, menzerInvestigationsModelingBRIR2009,amengualgariSpatialAnalysisAuralization2017}.

The listening experiment took place in the ITU-R BS.1116-compliant listening room (6.2m x 5.6m x 3.4m) at the University of Huddersfield's Applied Psychoacoustics Laboratory (APL). The HULTI-GEN Version 2 software provided the test interface \cite{johnsonHuddersfieldUniversalListening2020}. The study was structured around a multifactor design, focusing on two attributes (ATTR): Spatial Fidelity and Timbral Fidelity. The evaluation for each attribute was divided into six sessions, each dedicated to a different source position (POSITION). The average session duration was approximately 15 minutes. Furthermore, within each session, there were three specific trials (ITEM), each assessing a distinct type of program material. During each trial, participants rated 10 test conditions (SYSTEM), denoted as A through J (Table \ref{tab:systems}). Before the experiment, subjects were provided with a detailed instruction sheet. The purpose of the document was to familiarise subjects with the procedure and introduce them to the definitions of spatial and timbral fidelity and methodology.

\subsection{Measurement of Spatial Room Impulse Responses}
Impulse response measurements were conducted in an ITU-R BS.1116-compliant listening room (6.2m x 5.6m x 3.4m; RT 0.25s; NR 12) at the University of Huddersfield's Applied Psychoacoustics Laboratory (APL). The loudspeakers used for the measurements were Genelec 8040A, offering a free field frequency response within $\pm$ 2.0 dB across a range from 48 Hz to 20 kHz. The microphone systems used for the measurements were as follows.
\begin{itemize}
    \item{Neumann KU100 dummy head microphone}
    \item{mhAcoustics Eigenmike em32 (referred to as em32 hereon)}
    \item{A 6OM1 open microphone array, comprising six omnidirectional Line Audio OM1 microphones. These are arranged in a three-dimensional grid with each microphone pair spaced 100 mm apart along the X, Y, and Z axes, closely mimicking the GRAS 50VI intensity probe array as in \cite{tervoSpatialDecompositionMethod2013}}
\end{itemize}
Binaural impulse responses acquired using the KU100 were used to create reference stimuli for the listening tests. Impulse responses with the em32 were used for rendering stimuli for the SDM and HO-SIRR methods. The particular microphone system was chosen for its high number of capsules (32), which might enhance the direction of arrival (DOA) estimation \cite{tervoSpatialDecompositionMethod2013}. In contrast, the open mic array was used for SDM and chosen based on studies suggesting its optimal performance in terms of DOA error and perceptual quality \cite{amengualgariOptimizationsSpatialDecomposition2021,ahrensPerceptualEvaluationBinaural2019}. 

For the KU100 and 6OM1, the Merging Horus audio interface served as the AD/DA converter and microphone preamp. The measurements performed with Eigenmike em32 involved the use of the Eigenmike Microphone Interface Box (EMIB) as the recording device and Merging Horus as the playback device. To counter potential impulse response distortions due to clock mismatch between devices \cite{farinaAdvancementsImpulseResponse2007}, word clock was used for synchronization, with the Merging Horus device set as master.

The Exponential Sine Sweep (ESS) was used as an excitation signal as described in \cite{farinaAdvancementsImpulseResponse2007}, with the following parameters: a sample rate of 48 kHz, frequency range of 20 Hz to 20 kHz, sweep length of 20 seconds, and fade in/out of 10 ms.

The acoustic centers of the microphone systems were positioned at a height of 127.5 cm from the floor, aligning with the heights of the acoustic axes of the zero elevation loudspeakers, which form the zero elevation plane of the system. Although our measurements encompassed a complete Dolby 7.1.4 setup, conforming to the 4+7+0 loudspeaker layout as per \cite{itu-rbs.2051-3AdvancedSoundSystem2022}, the study primarily focused on a subset of these configurations, as detailed in Table \ref{tab:loudspeaker_positions}. For azimuth angle measurements, defined as the angle relative to the zero azimuth, each loudspeaker in the zero elevation level was placed at a distance of 2.00 m ($\pm 0.02\ \text{m}$) from the center of the microphone array. For the elevation angle measurements, the loudspeakers were placed at a distance of 1.92 m ($\pm 0.1\ \text{m}$) from the center of the microphone array.

\begin{table}[t]
  \centering
  \caption{Tested Loudspeaker Positions in the ITU-R 4+7+0 Layout}
  \begin{tabular}{lll}
  \hline
  \textbf{Position} & \textbf{Azimuth} & \textbf{Elevation} \\
  \hline
  Front center & $0 \degree$ & $0 \degree$ \\
  Front left & $30 \degree$ & $0 \degree$ \\
  Side left & $90 \degree$ & $0 \degree$ \\
  Back left & $135 \degree$ & $0 \degree$ \\
  Upper front left & $45 \degree$ & $45 \degree$ \\
  Upper back left & $135 \degree$ & $45 \degree$ \\
  \hline
  \end{tabular}
  \label{tab:loudspeaker_positions}
\end{table}

\subsection{Test Conditions and Variables}

In this study, SDM conditions were generated using the SDM Toolbox\footnote{SDM Toolbox (version 1.3001, Updated 22 Apr 2018).} \cite{tervoSDMToolbox2023}, while PIV-based DOA estimation algorithm was adapted from Zaunschirm et al.\cite{zaunschirmBinauralRenderingMeasured2020}. Given BSDM's growing relevance \cite{helmholzPredictionPerceivedRoom2022, bedernaPerceptualDetectionThresholds2023, mullerPerceptualDifferencesModifications2022, surduLIAudiovisualDataset2023,meyer-kahlenTwoDimensionalThresholdTest2023}, we included two conditions utilizing the available toolbox\footnote{BinauralSDM (version 0.5, commit 965da5c).} \cite{amengualgariBinauralSDM2023}.

HO-SIRR condition was rendered via its MATLAB implementation\footnote{HO-SIRR (commit 085f20d).} \cite{mccormackHOSIRR2023}. Both SDM utilizing PIV-based DOA analysis and HO-SIRR conditions, employed spherical harmonics from the Eigenmike em32, with SDM additionally using Line Audio OM1 as a pressure signal in one of the conditions.

Settings for SDM, HO-SIRR, and BSDM adhered to the recommended configurations in their respective toolboxes \cite{amengualgariBinauralSDM2023,tervoSDMToolbox2023,mccormackHOSIRR2023}. The exception was the window size, which was standardized at 64 samples across all frameworks, following Amengual Gari et al.'s recommendations \cite{amengualgariOptimizationsSpatialDecomposition2021}. Additionally, post-equalization \cite{tervoSpatialAnalysisSynthesis2015} was disabled in the SDM Toolbox.

Both BSDM and PIV-based methods applied band-limited DOA estimation (200 Hz-2400 Hz), aligned with the spatial aliasing frequency of the 6OM1 array. BSDM's mixing time was set at 38 ms, based on ISM simulations of our study room.

We ensured that the anchor demonstrated impairments in both spatial and timbral fidelity. Specifically, for azimuthal sources at zero elevation positions, the anchor employed a KU100 BRIR that was offset by an additional 60 to 90 degrees from the source position being evaluated. For sources at elevated positions, given the limited number of measured positions, a BRIR corresponding to a diametrically opposite position was chosen, 180 degrees from the source position being evaluated. To introduce a timbral fidelity impairment, we applied a 3.5kHz low-pass filter \cite{itu-rbs.1534-3MethodSubjectiveAssessment2014}.

\subsection{Synthesis of Binaural Room Impulse Responses}

To facilitate a subjective experiment, we employed binaural rendering techniques for SDM, BSDM, and HO-SIRR to incorporate the KU100 BRIR as a ground truth reference. Consequently, we employed a dataset of 2702 KU100 head-related impulse responses (HRIRs) sampled on the Lebedev grid \cite{bernschutzSphericalFarField2013}. The synthesis process entailed rendering virtual loudspeaker signals at points on the Lebedev grid using the evaluated systems. For SDM, this was achieved using K-nearest neighbor allocation, while HO-SIRR used VBAP. The final stage involved convolving these virtual loudspeaker signals with corresponding HRTFs from the SOFA file \cite{majdakSpatiallyOrientedFormat2022} and summing them to produce the final BRIRs.

\subsection{Programme Material}
Accurate binaural sound-field reproduction requires anechoic audio, free from original recording environment effects. This is achieved by convolving the anechoic material with binaural room impulse responses (BRIR), simulating room acoustics. The anechoic samples employed in our study are as follows: "Bongo" (Track 26, 12.780s-–20.330s) and "Danish Speech" (Track 9, 0.590s-–9.240s) from Bang \& Olufsen's "Music for Archimedes" \cite{bang&olufsenMusicArchimedes1992,hansenMakingRecordingsSimulation1991}, and Handel/Harty's "No.6 Water Music Suite" (Track 9, 0.480s-–8.380s) from Denon's anechoic orchestral collection \cite{variousartistsAnechoicOrchestralMusic1988}.

\subsection{Reproduction Configuration and Calibration}

The study employed a reproduction system that included AKG K702 headphones and a Merging Technologies Horus audio interface. These headphones were fitted with an inverse filter originally developed for Lee et al.'s research \cite{leeSpatialTimbralFidelities2019}, designed to correct spectral coloration and replicate the response in a KU100 dummy head's ear. The filter's creation involved placing AKG K702 headphones on a KU100 dummy head and conducting five measurements for each ear in both left/right and right/left positions, then repeating the process with another pair of headphones, resulting in 40 impulse responses in total. The filter design followed the high-pass-regularized least-mean-square (LMS) inversion approach \cite{kirkebyDigitalFilterDesign1999}, identified as a perceptually sound inversion algorithm \cite{scharerEvaluationEqualizationMethods2009}.

Test stimuli were uniformly normalized to -26 LUFS and the headphone amp gain was set uniformly throughout the experiment. LAeq loudness for KU100 BRIR at +30\degree\ azimuth, measured via miniDSP EARS, averaged 70.7dB (Bongo), 77.5dB (Speech), and 73.5dB (Orchestra).

\subsection{Test Subjects}
In total, 14 participants took part in the study. These participants ranged in age from 20 to 47 years. They were a mixture of staff members, postgraduate students, and undergraduate researchers affiliated with the Applied Psychoacoustics Laboratory (APL) at the University of Huddersfield.  All participants claimed to have normal hearing abilities and had previously taken part in listening experiments. Prior to the formal listening test, the participants were asked to complete a short 10-minute familiarization phase for each attribute (ATTR).

\subsection{Objective Metrics}

The present study considered four objective metrics: Interaural Level Difference ($\mathrm{ILD}$), Interaural Time Difference ($\mathrm{ITD}$), reverberation time ($\mathrm{T30_{mid}}$), early Interaural Cross-Correlation Coefficient ($\mathrm{IACC_{E3}}$) and late IACC ($\mathrm{IACC_{L3}}$). To evaluate accuracy, the Mean Absolute Error (MAE) quantifies the error's magnitude, while the Mean Signed Difference (MSD) detects systematic biases, providing insights into the extent and direction of the system's performance deviations.

Each BRIR was energy-normalized to allow fair comparison. This involved identifying the onset of direct sound, indicated by the earliest sound arrival in either channel and calculating the Root Mean Square (RMS) value over a 2.5 ms segment starting from this point \cite{zahorikDirecttoreverberantEnergyRatio2002}. This RMS value, representing the direct sound's energy, was then used to normalize the entire BRIR, thus minimizing gain differences and establishing a consistent baseline for objective analysis.

\subsubsection{Interaural Level Difference (ILD)}
The ILD is a major cue for horizontal sound localization, accentuated by head shadowing when sources are off-center, with the Just Noticeable Difference (JND) ranging from 1 to 2 dB \cite{blauertSpatialHearingPsychophysics1997}. Accounting for the precedence effect, ILD was derived from the first 2.5 ms of BRIRs over 39 equivalent rectangular bandwidth (ERB) bands and averaged as follows:
\begin{equation}
  \mathrm{ILD}_{\text{avg}} = \frac{1}{N} \sum_{f=1}^{N} 20 \log_{10}\left(\frac{\hat{y}_{L}(f)}{\hat{y}_{R}(f)}\right)
\end{equation}
where $\mathrm{ILD}_{\text{avg}}$ is the average ILD over $N$ ERB bands, with $\hat{y}_{L}(f)$ and $\hat{y}_{R}(f)$ as the RMS values for left and right channels at each frequency band $f$. 

\subsubsection{Interaural Time Difference (ITD)}

The ITD is the second of two auditory cues, critical for lateral sound localization, with JND of about 40 $\mu s$ for frontal sources and approximately 100 $\mu s$ for lateral sources \cite{andreopoulouIdentificationPerceptuallyRelevant2017}. The ITD can extend up to 700 $\mu s$ for azimuth angles up to 90 degrees. ITD estimation is based on the peak time lag of the interaural cross-correlation function:
\begin{equation}
\text{ITD} = \arg \max_{-1\ \text{ms} < \tau < 1\ \text{ms}} (|\text{IACF}(\tau)|)
\end{equation}

\subsubsection{Reverberberation Time ($\text{RT}_{60}$)}

The $\text{RT}_{60}$, was calculated in octave bands in accordance with the standards outlined in \cite{iso3382-1AcousticsMeasurementRoom2009}. The value presented was derived based on a 30 dB evaluation range and subsequently averaged across the 500 Hz and 1 kHz octave bands, resulting in $\text{T30}_{mid}$. The JND for reverberation time is established at 5\% of the $\text{RT}_{60}$ \cite{iso3382-1AcousticsMeasurementRoom2009}. With a 0.25-second reverberation time in the auralized room, the JND in the present scenario is roughly 12.5 ms.

\subsubsection{Interaural Cross Correlation Coefficient (IACC)}
The IACC is a commonly used metric of spatial impression (SI) in concert halls, with a JND of 0.075 \cite{iso3382-1AcousticsMeasurementRoom2009}. The Apparent Source Width (ASW) and Listener Envelopment (LEV) are best estimated by averaging the IACC across 500 Hz, 1 kHz, and 2 kHz octave bands, resulting in $\text{IACC}_{E3}$ and $\text{IACC}_{L3}$ metrics \cite{hidakaInterauralCrossCorrelation1995}. In the present study, we calculated $\text{1-IACC}_{E3}$ and $\text{1-IACC}_{L3}$, which are positively correlated with ASW and LEV. The IACC is computed as follows:
\begin{equation}
\text{IACC} = \max_{-1\ \text{ms} < \tau < 1\ \text{ms}} |\text{IACF}(\tau)|
\end{equation}

\section{Subjective Evaluation Results} \label{sec:subjective}

\begin{table}[b]
  \centering
  \caption{Results of the aggregated Friedman test for different factors and attributes}
  \begin{tabular}{llllll}
  \hline
  \textbf{ATTR} & \textbf{Factor} & \textbf{Test Statistic ($\chi^2$)} & \textbf{DF} & \textbf{p-value} & \textbf{sig.} \\
  \hline
  Spatial & SYSTEM & 101 & 9 & $<$0.0001 & **** \\
  Spatial & ITEM & 2.43 & 2 & 0.297 & ns \\
  Spatial & POSITION & 21.4 & 5 & 0.0007 & *** \\
  Timbral & SYSTEM & 106 & 9 & $<$0.0001 & **** \\
  Timbral & ITEM & 0.531 & 2 & 0.767 & ns \\
  Timbral & POSITION & 11.5 & 5 & 0.0417 & * \\
  \hline
  \end{tabular}
  \label{tab:friedman}
\end{table}

\begin{figure}[!b]
  \centering
  \includegraphics[width=0.85\columnwidth]{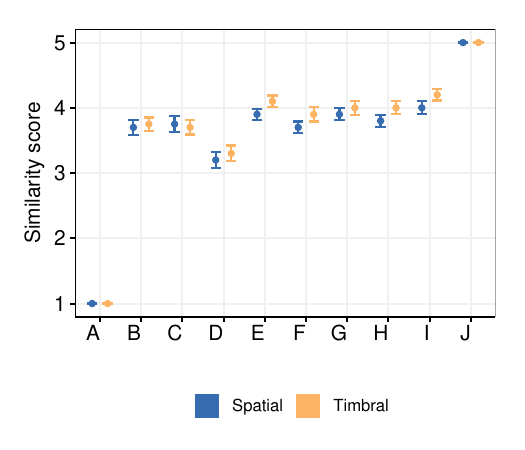}
  \caption{Subjective evaluation results showing median spatial and timbral fidelity scores on similarity scale (1-5), for various systems (SYSTEM). Scores are compiled from multiple source positions (POSITION) and program materials (ITEM), with 95\% non-parametric confidence intervals.}
  \label{fig:overall_result}
\end{figure}

Following ITU guidelines \cite{itu-rbs.1534-3MethodSubjectiveAssessment2014}, a post-screening process was implemented to identify and potentially exclude assessors who had rated the hidden reference below 4.5 (90\% of the 5.0 maximum score) for over 15\% of test items. This was done to ensure that only reliable subjects were considered. Ultimately, none of the assessors met these exclusion criteria.

To verify the normality of the ratings given to each system for each combination of factors, the data were grouped by SYSTEM, ATTR, ITEM, and POSITION. Consistently, the ratings assigned to systems 'A' (Reference) and 'J' (Anchor) were found not to follow a normal distribution. Excluding these, non-normal distributions were found in 6.25\% of cases for the spatial fidelity and approximately 6.94\% for the timbral fidelity attribute. Considering these results, we opted for a non-parametric statistical approach in subsequent analyses.

\begin{table}[b]
  \caption{Systems under test}
  \label{tab:systems}
  \centering
  \begin{tabular}{c p{5cm}}
    \hline
    \textbf{Label} & \textbf{Test system (SYSTEM)} \\
    \hline
    A & Anchor \\
    B & BSDM 6OM1 Omni \\
    C & BSDM em32 Omni \\
    D & SDM em32 \\
    E & SDM em32 Omni \\
    F & SDM 6OM1 Omni \\
    G & HO-SIRR diffuse \\
    H & SDM PIV \\
    I & SDM PIV Omni \\
    J & KU100 (Reference) \\
    \hline
  \end{tabular}
\end{table}

\begin{figure*}[!h]
  \centering
  \includegraphics[width=\textwidth]{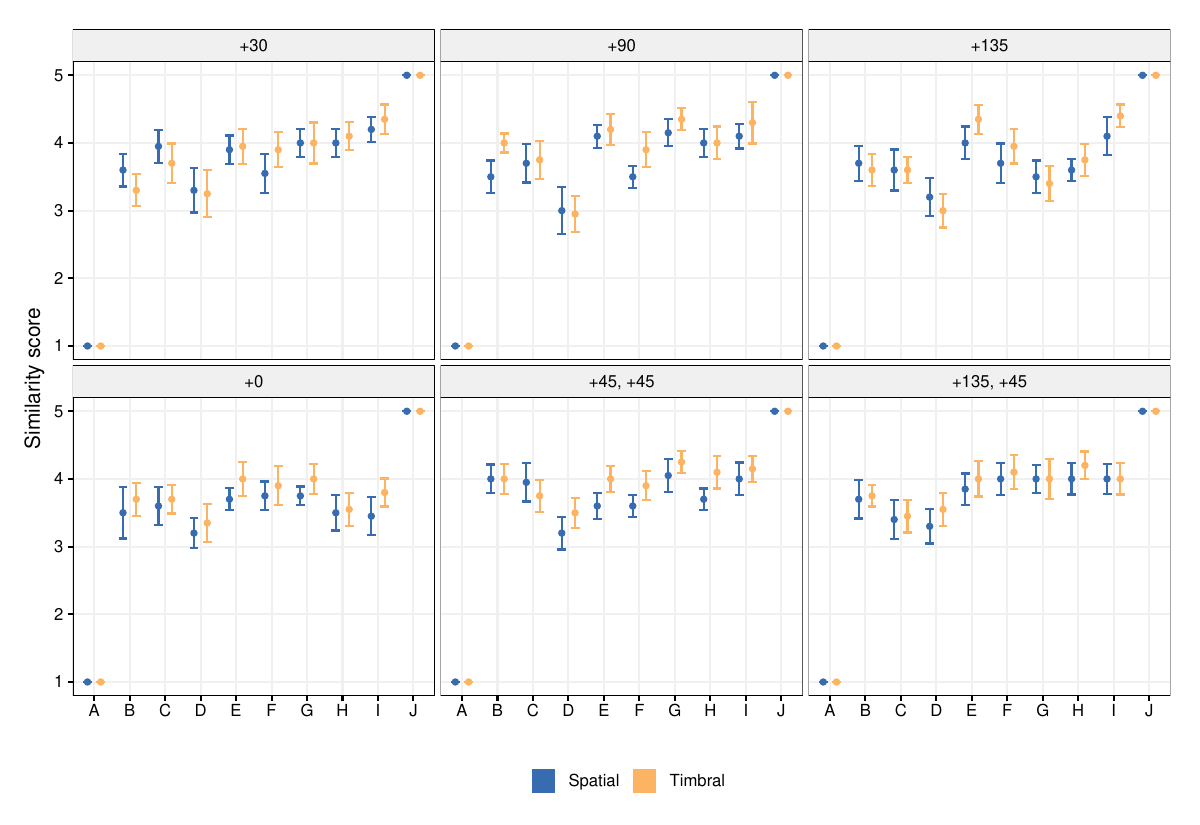}
  \caption{Subjective evaluation results showing median for spatial and timbral fidelity scores on similarity scale (1-5), across different systems (SYSTEM) and source positions (POSITION), aggregated over all program materials (ITEM). The graph includes 95\% non-parametric confidence intervals.}
  \label{fig:result_pos}
\end{figure*}

\begin{figure*}[!h]
  \centering
  \includegraphics[width=\textwidth]{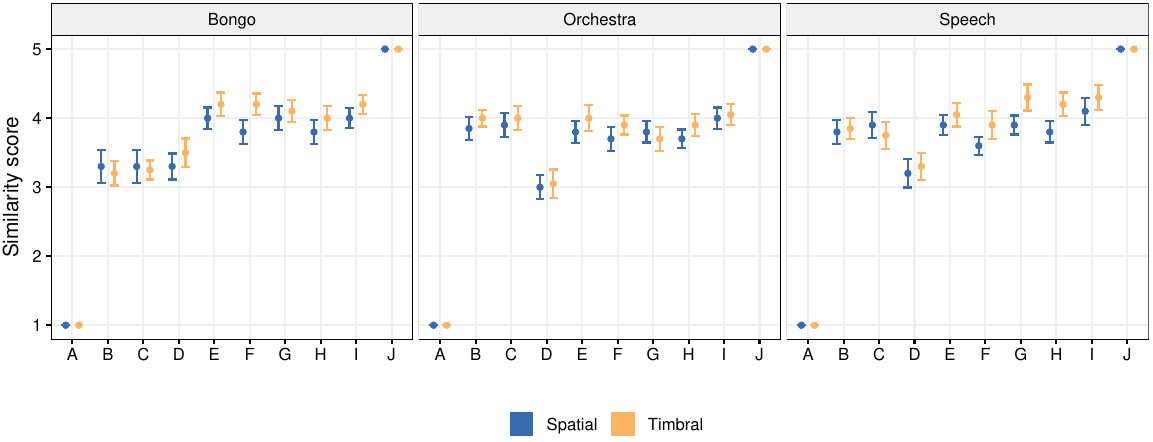}
  \caption{Subjective evaluation results showing median for spatial and timbral fidelity scores on similarity scale (1-5), across different systems (SYSTEM) and program materials (ITEM), aggregated over all source positions (POSITION). The graph includes 95\% non-parametric confidence intervals.}
  \label{fig:result_item}
\end{figure*}

Friedman's tests were conducted to examine the main effects of various factors---SYSTEM, ITEM, POSITION---on both spatial and timbral fidelity attributes. The detailed statistical outcomes of these tests, including test statistics and significance levels, are summarized in Table~\ref{tab:friedman}.

These results underscore a significant influence of the SYSTEM on both spatial and timbral attributes, indicating variations in the capability of the rendering methods to replicate spatial and timbral fidelity. Interestingly, the specific program material (ITEM) did not have a significant impact on participant ratings, while the source position (POSITION) demonstrated a noticeable effect, especially in spatial fidelity.

First, we present the aggregated results across various ITEMs and POSITIONs in Figure~\ref{fig:overall_result}, giving an overview of the overall system performance. Given the significant effect of POSITION and the non-significant effect of ITEM from our Friedman test findings, we further delved into Position-Based Performance (Figure~\ref{fig:result_pos}) and potential subtle influences of Program Material (Figure~\ref{fig:result_item}). Disparities between rendering systems were analyzed using the Wilcoxon signed-rank test with Holm-Bonferroni correction \cite{holmSimpleSequentiallyRejective1979}. Additionally, we calculated the effect size ($r$) using the method described in \cite{fieldDiscoveringStatisticsUsing2012}, which involves converting the p-value to a z-score and then to $r$, based on the total number of observations (N).

Figure \ref{fig:overall_result} shows the ratings for spatial and timbral Fidelity of various systems. Low anchor (A), consistently scored at the lower end of the scale with a median of 1.0, indicating 'Extremely Different'. In contrast, KU100 (J) achieved a median score of 5.0 ('Same') implying that subjects had no problem in identifying the hidden reference. Systems BSDM 6OM1 Omni (B) to SDM PIV Omni (I) varied, with spatial fidelity ratings mainly between 3 and 4 and timbral fidelity also clustering between 3 and 4.2.

Looking more closely at spatial fidelity, SDM em32 Omni (E), HO-SIRR (G), and SDM PIV Omni (I) displayed comparable performance (medians between 3.9 and 4). However, SDM PIV (H) with a median of 3.8 was slightly inferior to both SDM em32 Omni (E) and HO-SIRR (G). SDM 6OM1 Omni (F) trailed behind these three systems. Additionally, BSDM 6OM1 Omni (B), BSDM em32 Omni (C), and SDM 6OM1 Omni (F) were grouped closely together (medians between 3.7 to 3.8), though BSDM em32 Omni (C) was outperformed by both HO-SIRR (G) and SDM PIV (H). Notably, SDM em32 (D) performance was significantly lower than the rest (median 3.2).

In terms of timbral fidelity, SDM em32 Omni (E) and SDM PIV Omni (I) were almost indistinguishable (medians between 4.1 and 4.2), with SDM em32 Omni (E) outperforming HO-SIRR (G) and SDM PIV (H). SDM 6OM1 Omni (F), HO-SIRR (G), and SDM PIV (H) were on par with each other (medians between 3.9 and 4), whereas BSDM 6OM1 Omni (B) and BSDM em32 Omni (C), while similar to each other (medians 3.7), lagged behind the preceding group, and even more so when compared to SDM em32 Omni (E) and SDM PIV Omni (I). Again, SDM em32 (D) scored substantially lower than BSDM 6OM1 Omni (B) and BSDM em32 Omni (C).

\subsection{Dedicated center Omnidirectional Microphone}

The impact of using a dedicated center omnidirectional microphone can be analyzed by comparing SDM em32 Omni (E) and SDM em32 (D)---utilizing TDOA-based DOA, as well as SDM PIV Omni (I) and SDM PIV (H)---utilizing PIV-based DOA. These conditions used SRIR from em32 (and obtained SPH), with and without center omnidirectional microphone.

The dedicated omnidirectional microphone (SDM em32 Omni, E) notably enhanced the em32 array's performance in the context of SDM's TDOA-based DOA variant (SDM em32, D), with significant improvements in overall results for both spatial and timbral fidelity, exhibiting a large effect size ($p<0.001$, $r>0.7$). The same trend was also observed for individual positions and stimuli, however, for certain positions with medium effect size, for instance, +45\degree, +45\degree\ (Spatial) and +135\degree, +45\degree\ (Timbral) with $p < 0.005$ and $r = 0.4$.

Improvements were also noticeable with a dedicated center microphone in SDM's PIV-based DOA estimation variant, but they were less pronounced. SDM PIV (H) used a zeroth-order eigenbeam, and SDM PIV Omni (I), which utilized a dedicated omni microphone, showed significant enhancement in spatial and timbral fidelity, albeit with a small effect size ($p<0.009$, $r<0.3$). Interestingly, for the +135\degree\ source position, the improvement was significant with a large effect size ($p<0.001$, $r=0.61$). Across most source positions, the improvement for both attributes was observed with a small effect size ($p<0.04$, $r<0.26$), except for Bongo and Speech in spatial fidelity.

\subsection{Microphone Array and DOA Estimation Method}
This research evaluated TDOA and PIV-based DOA estimation methods in SDM for spatial data capture and auralization. It focused on TDOA-based DOA using em32 and 6OM1 arrays, represented by SDM em32 Omni (E) and SDM 6OM1 Omni (F), and PIV-based DOA with first-order SPH, denoted as SDM PIV Omni (I), all utilizing a central pressure signal from the array.

As depicted in Figure \ref{fig:overall_result}, our results indicate that SDM em32 Omni (E), which relies on TDOA-based DOA estimation, exhibited performance on par with SDM PIV Omni (I)---utilizing a PIV-based DOA estimation---when the identical omnidirectional pressure signals were used. The comparison revealed no substantial differences in either spatial or timbral fidelity. Conversely, SDM 6OM1 Omni (F)---also based on TDOA but incorporating an array of only six microphones, was found to perform significantly worse than SDM em32 Omni (E) with a small effect size ($p < 0.001$, $r < 0.26$) and SDM PIV Omni (I) with medium effect size ($p < 0.001$, $0.3 < r < 0.36$) for both spatial and timbral fidelity.

Upon examining the specific source positions (Figure \ref{fig:result_pos}), we observed a similar pattern. SDM 6OM1 Omni (F) generally underperformed compared to SDM em32 Omni (E) and SDM PIV Omni (I). This was particularly pronounced for source positions at +30\degree\ and +90\degree, where SDM 6OM1 Omni (F) demonstrated significant deviation from SDM em32 Omni (E) ($p < 0.007$, medium and large effect size: $0.41 < r < 0.61$) and SDM PIV Omni (I) ($p < 0.001$, large effect: $0.66 < r < 0.69$) concerning spatial fidelity. On the other hand, it was observed that SDM PIV Omni (I) outperformed SDM em32 Omni (E) at +30\degree\ and +45\degree, +45\degree\ source positions with medium effect size ($p < 0.013$, $0.38 < r < 0.46$) in terms of spatial fidelity. Regarding timbral fidelity, SDM em32 Omni (E) and SDM PIV Omni (I) notably outperformed SDM 6OM1 Omni (F) at the +135\degree\ position ($p < 0.001$, large effect: $0.5 < r < 0.54$), with SDM PIV Omni (I) also showing superior performance at the +30\degree\ and +90\degree\ source positions with medium effect size ($p = 0.011$, r = 0.39). 

An analysis of stimulus-dependent performance revealed that there were no significant differences between SDM em32 Omni (E) and SDM PIV Omni (I), except in two specific scenarios. SDM PIV Omni (I) exhibited superior spatial fidelity over SDM em32 Omni (E) for Orchestra stimuli, with a small effect size ($p < 0.045$, $r = 0.22$). For Speech stimuli, SDM PIV Omni (I) also showed greater timbral fidelity than SDM em32 Omni (E), again with a small effect size ($p < 0.028$, $r = 0.24$). In contrast, SDM 6OM1 Omni (F) underperformed in both spatial and timbral fidelity for Speech compared to SDM em32 Omni (E) and SDM PIV Omni (I), with a medium effect size ($p < 0.005$, $0.3 < r < 0.36$). However, the exception was the comparison of timbral fidelity between SDM PIV Omni (I) and SDM 6OM1 Omni (F), where SDM PIV Omni (I) demonstrated a large effect size ($p < 0.001$, $r = 0.57$). Additionally, for Orchestra stimuli, SDM PIV Omni (I) outperformed SDM 6OM1 Omni (F) in both spatial and timbral fidelity with medium ($p < 0.001$, $r = 0.36$) and small ($p < 0.022$, $r = 0.24$) effect sizes, respectively.

\subsection{Spatial Encoding Systems}

This section analyzes the impact of different rendering systems on spatial encoding efficacy, comparing SDM's performance in SDM em32 (D), SDM em32 Omni (E), SDM 6OM1 Omni (F), SDM PIV (H), and SDM PIV Omni (I), with  HO-SIRR (G), and BSDM's optimizations in BSDM 6OM1 Omni (B) and BSDM em32 Omni (C).

It is noteworthy that no enhancement in spatial fidelity was observed between SDM 6OM1 Omni (F) and BSDM 6OM1 Omni (B). Intriguingly, SDM 6OM1 Omni (F) generally outperformed BSDM 6OM1 Omni (B) in terms of timbral fidelity, albeit with a modest effect size ($p=0.001$, $r=0.2$), considering the overall results. A similar trend was evident for BSDM em32 Omni (C) and SDM em32 Omni (E), indicating that the optimizations employed by BSDM frequently led to degradation of these attributes, exhibiting small ($p<0.001$, $r=0.25$) and medium ($p<0.001$, $r=0.49$) effect sizes for spatial and timbral fidelity, respectively.

Looking at how performance varies with position, BSDM 6OM1 Omni (B) generally matched the performance of SDM 6OM1 Omni (F) in terms of spatial and timbral fidelity, with the exception of the +30\degree\ position. At this position, SDM 6OM1 Omni (F) outperformed BSDM 6OM1 Omni (B) with a large effect size ($p < 0.001$, $r = 0.54$) in terms of timbral fidelity. Similarly, at +30\degree\ BSDM 6OM1 Omni (B) underperformed compared to systems SDM em32 Omni (E) to SDM PIV Omni (I) with a large effect size ($p < 0.001$, $r > 0.52$) in terms of timbral fidelity. In terms of spatial fidelity at +30\degree, BSDM 6OM1 Omni (B) did not show a significant difference when compared to SDM 6OM1 Omni (F), yet HO-SIRR (G), SDM PIV (H), and SDM PIV Omni (I) outperformed BSDM 6OM1 Omni (B) with a large effect at +90\degree\ ($p < 0.001$, $r > 0.5$). Additionally, at the +135\degree, +45\degree\ position, BSDM 6OM1 Omni (B) underperformed with a medium effect size ($p < 0.016$, $r = 0.37$) in terms of both spatial and timbral fidelity compared to SDM 6OM1 Omni (F).

\begin{figure*}[t]
  \centering
  \subfloat[Spatial Fidelity]{%
    \includegraphics[width=0.49\textwidth]{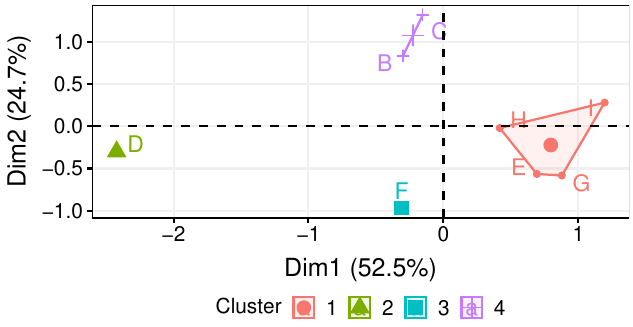}%
    \label{fig:spatial-overall}%
  }
  \hfill
  \subfloat[Timbral Fidelity]{%
    \includegraphics[width=0.49\textwidth]{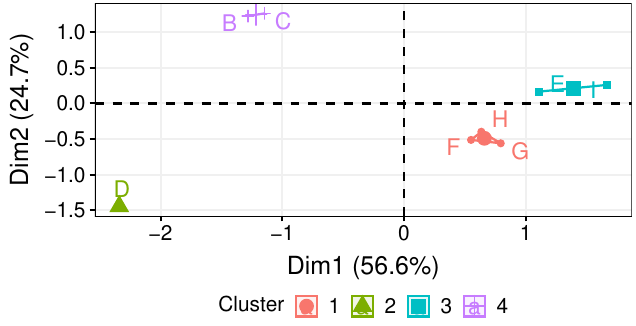}%
    \label{fig:spatial-overall}%
  }
  \caption{  Principal Component Analysis (PCA) of spatial and timbral fidelity scores for evaluated systems (SYSTEM), considering the median spatial and timbral fidelity scores across different program materials (ITEM) and source positions (SYSTEM). Systems are clustered in a two-dimensional space by the first two principal components, highlighting the similarities in their fidelity scores.}
  \label{fig:pca-pos}
\end{figure*}

Similar observations were made for BSDM em32 Omni (C) and SDM em32 Omni (E). Both systems performed similarly for most positions, except at the +90\degree\ and +135\degree, +45\degree\ source positions in terms of both spatial and timbral fidelity. At these positions, SDM em32 Omni (E) outperformed BSDM em32 Omni (C) with large ($p < 0.001$, $0.51 < r < 0.65$) and medium ($p < 0.002$, $0.44 < r < 0.48$) effect sizes in terms of spatial fidelity. However, in terms of timbral fidelity, the differences were less pronounced, with the main differences observed at the +135\degree, +45\degree\ source position. Moreover, BSDM em32 Omni (C) underperformed for spatial fidelity when compared with systems SDM em32 Omni (E) to SDM PIV Omni (I) at +135\degree, +45\degree\, and +90\degree\ with large ($p < 0.001$, $r > 0.5$) and medium effect sizes ($p < 0.02$, $0.35 < r < 0.45$).

Examining the results for individual stimuli reveals an interesting pattern regarding spatial fidelity. BSDM 6OM1 Omni (B) and BSDM em32 Omni (C) performed comparably to systems utilizing SDM (E, F, H, and I) and HO-SIRR (G) for Orchestra and Speech stimuli in terms of spatial fidelity. However, their performance significantly deteriorated for the Bongo stimulus, with a large effect size ($p < 0.001$, $r > 0.5$). A similar trend was observed for timbral fidelity, though BSDM 6OM1 Omni (B) and BSDM em32 Omni (C) occasionally underperformed compared to systems SDM em32 Omni (E), SDM 6OM1 Omni (F), SDM PIV (H), and SDM PIV Omni (I) for Orchestra and Speech stimuli with small to large effect sizes.

While we previously concluded that systems utilizing SDM--SDM em32 Omni (E) and SDM PIV Omni (I) tend to perform better than SDM 6OM1 Omni (F), this observation suggests the impact of the microphone array used. When examining the performance differences between SDM and HO-SIRR (G) in terms of spatial fidelity, we observe no significant difference between HO-SIRR (G), SDM em32 Omni (E), SDM PIV (H), and SDM PIV Omni (I). However, HO-SIRR (G) shows significantly better performance compared to SDM 6OM1 Omni (F) with a small effect size ($p < 0.009$, $r = 0.16$). On the other hand, it performed worse than SDM em32 Omni (E) and SDM PIV Omni (I) in terms of timbral fidelity, but with a small effect size ($p < 0.018$, $0.15 < r < 0.27$). Interestingly, HO-SIRR (G) and H (SDM PIV) did not demonstrate a significant difference in terms of spatial and timbral fidelity, which may imply that the use of the zeroth-order eigenbeam as the pressure signal in SDM PIV (H) and HO-SIRR (G) is a contributing factor.

The results for individual source positions implied that HO-SIRR (G) performed similarly to SDM em32 Omni (E) and SDM PIV Omni (I) in most of the evaluated source positions, with the occasional exceptions when it underperformed, for instance at +135\degree\ source position with large effect size ($p = 0.001$, $r > 0.5$). Conversely, at +45\degree, +45\degree\ source position, HO-SIRR (G) outperformed SDM em32 Omni (E) with a medium effect size for spatial fidelity  ($p = 0.024$, $r > 0.34$). Interestingly, the performance of HO-SIRR (G) mirrored that of SDM PIV (H) in spatial and timbral fidelity for individual stimuli, with only occasional differences. For instance, regarding timbral fidelity, HO-SIRR (G) was outperformed by SDM em32 Omni (E) and SDM PIV Omni (I) for the Orchestra stimulus with a medium effect size ($p < 0.004$, $r > 0.31$).

\subsection{Principal Component Analysis}

In the spatial analysis and auralization of a critical listening room, SDM and HO-SIRR systems exhibited similar performance, particularly in terms of spatial fidelity. Systems such as SDM em32 Omni (E), HO-SIRR (G), SDM PIV Omni (I), and SDM PIV (H) demonstrated comparable results, though with some variations. With respect to timbral fidelity, SDM em32 Omni (E) and SDM PIV Omni (I), which utilize TDOA-based analysis with Eigenmike em32 and PIV-based analysis with a dedicated pressure signal, closely aligned with the reference.

This alignment is supported by Principal Component Analysis (PCA), as shown in Figure \ref{fig:pca-pos}. The PCA clusters these systems together based on their median spatial and timbral fidelity scores across all positions and stimuli. The analysis revealed that the first two principal components accounted for about 77\% of the variance in spatial fidelity and 80\% in timbral fidelity.

\subsection{Correlation between Spatial and Timbral fidelity}

\begin{figure}[t]
  \centering
  \includegraphics[width=\columnwidth]{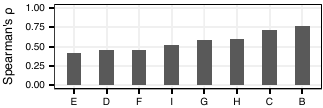}
  \caption{Correlation between median scores of spatial and timbral fidelity across all source positions and stimuli.}
  \label{fig:correlation}
\end{figure}

To explore the link between spatial and timbral fidelity, we calculated Spearman's correlation based on median scores across all positions and stimuli, as depicted in Figure \ref{fig:correlation}.

The weakest correlations between spatial and timbral fidelity were observed in SDM em32 Omni (E, $\rho = 0.41$), SDM em32 (D), and SDM 6OM1 Omni (F, $\rho = 0.45$), followed by SDM PIV Omni (I, $\rho = 0.52$). A significant increase in correlation was noted for HO-SIRR (G, $\rho = 0.59$) and SDM PIV (H, $\rho = 0.6$), with BSDM em32 Omni (C, $\rho = 0.72$) and BSDM 6OM1 Omni (B, $\rho = 0.77$) exhibiting the strongest positive correlations.

The strong correlation in BSDM-rendered conditions (B and C) indicates a uniform impact of artifacts on both spatial and timbral fidelity, possibly due to reverb equalization effects. Conversely, the pronounced correlation between spatial and timbral fidelity in HO-SIRR (G) and SDM PIV (H) implies an effect resulting from the use of a zeroth-order spherical harmonic as a pressure signal within these systems.

\section{Objective Evaluation Results} \label{sec:objective}

\begin{figure*}[hb]
  \centering
  \subfloat[ILD computed from BRIRs across 39 ERB bands, averaged above and below 1.5 kHz as detailed in \cite{lee3DMicrophoneArray2021}.]{%
    \includegraphics[width=0.49\textwidth]{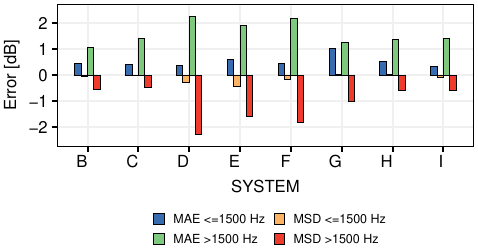}%
    \label{fig:ildmaemsd}%
  }
  \hfill
  \subfloat[Interaural cross-correlation coefficients: Early ($\text{1-IACC}_{E3}$) and Late ($\text{1-IACC}_{L3}$).]{%
    \includegraphics[width=0.49\textwidth]{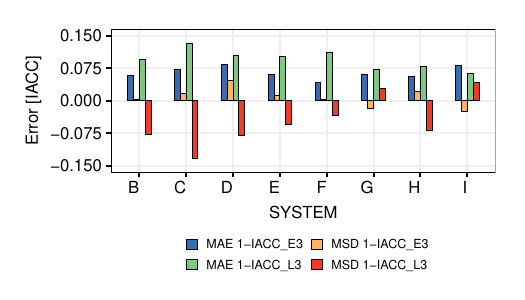}%
    \label{fig:iaccmaemsd}%
  }

  \subfloat[ITD calculated using maximum interaural cross-correlation (MaxIACCr) method.]{%
    \includegraphics[width=0.49\textwidth]{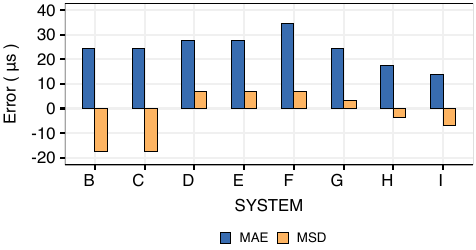}%
    \label{fig:itdmaemsd}%
  }
  \hfill
  \subfloat[Mid-frequency reverberation time $\text{T30}_{mid}$ as defined in \cite{iso3382-1AcousticsMeasurementRoom2009}.]{%
    \includegraphics[width=0.49\textwidth]{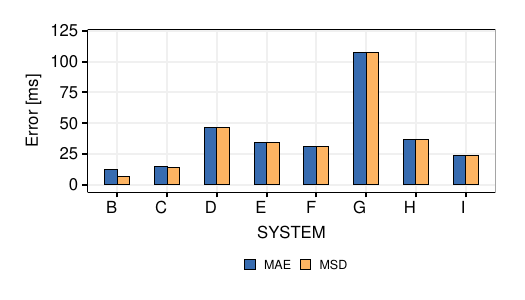}%
    \label{fig:t30maemsd}%
  }
  \caption{Objective metrics presented using mean absolute error (MAE) and mean signed difference (MSD) using the reference KU100 (J) as the ground truth.}
  \label{fig:mainlabel}
\end{figure*}

\subsection{Interaural Level Difference (ILD)}

Figure \ref{fig:ildmaemsd} illustrates the mean absolute error (MAE) and mean signed difference (MSD) for ILD across systems (B through I), categorized by the ERB bands above and below 1500 Hz used for averaging. At higher frequencies, SDM em32 (D), SDM em32 Omni (E), and SDM 6OM1 Omni (F) showed notable deviation with high MAEs and negative MSDs, underscoring a tendency to underestimate ILDs beyond the established JND \cite{blauertSpatialHearingPsychophysics1997}. HO-SIRR (G) also tended to underestimate ILD but to a lesser extent. In contrast, lower frequencies exhibited reduced MAEs for most systems, except HO-SIRR (G), and MSDs near zero, indicating minimal bias except for the consistent underestimation by SDM em32 (D), SDM em32 Omni (E), and SDM 6OM1 Omni (F). Overall, all systems faced more difficulty with higher frequency ILDs, implying challenges in capturing spatial cues accurately. However, performance improved at frequencies below 1500 Hz, with SDM em32 (D) being the exception.

\subsection{Interaural Time Difference (ITD)}

Figure \ref{fig:itdmaemsd} demonstrates ITD accuracy with MAE and MSD values, indicating that BSDM 6OM1 Omni (B) and BSDM em32 Omni (C) consistently underestimate ITD, whereas SDM em32 (D), SDM em32 Omni (E), and SDM 6OM1 Omni (F) overestimate, particularly SDM 6OM1 Omni (F) with the highest MAE of 34.7 $\mu s$. HO-SIRR (G) has minor overestimations, while SDM PIV (H) and SDM PIV Omni (I) maintain the highest accuracy with slight underestimation tendencies. Given the JND for ITD ranges from 40 $\mu s$ frontally to 100 $\mu s$ laterally \cite{andreopoulouIdentificationPerceptuallyRelevant2017}, most systems' deviations fall within perceptually insignificant limits. Notably, there is an inconsistency in MSD and MAE magnitudes and a suggested ILD-ITD trade-off in systems SDM em32 (D) through HO-SIRR (G).

\subsection{Reverberberation Time}

Figure \ref{fig:t30maemsd} shows that all systems tend to overestimate reverberation time, as indicated by positive MSD values. Most systems' errors exceed the JND of 12.5 ms, which may affect perception, except for BSDM 6OM1 Omni (B) and BSDM em32 Omni (C) which maintain minimal perceptible errors with MAEs of 12.4 and 14.7 ms. These results align with Amengual et al.'s findings, confirming that RTMod maintains synthesized BRIR's reverberation time within the JND \cite{amengualgariOptimizationsSpatialDecomposition2021}. On the other hand, HO-SIRR (G), with an MAE of 108 ms, significantly overshoots the JND, hinting at a noticeable impact on spatial fidelity.

\subsection{Interaural Cross Correlation Coefficient (IACC)}

Figure \ref{fig:iaccmaemsd} presents MAE and MSD for systems using $\text{1-IACC}_{E3}$ and $\text{1-IACC}_{L3}$. Evaluated systems vary in accuracy, with $\text{1-IACC}_{E3}$ MAE ranging from 0.0427 to 0.0836 and $\text{1-IACC}_{L3}$ MAE from 0.0637 to 0.134. Errors for $\text{1-IACC}_{E3}$ mostly fall within the JND of 0.075, implying that the errors are unlikely to be perceptible. Low MSDs suggest non-systematic errors, except for SDM em32 (D) and SDM PIV Omni (I), which show a bias in $\text{1-IACC}_{E3}$ estimates, potentially affecting ASW.

In contrast, MAEs for $\text{1-IACC}_{L3}$ surpass the JND of 0.075 for all but HO-SIRR (G) and SDM PIV Omni (I), with several systems showing negative MSDs, hinting at consistent underestimation and possible LEV impact. HO-SIRR (G) and SDM PIV Omni (I), however, remain within JND bounds for both MAE and MSD, suggesting minimal LEV alteration.

\section{Discussion} \label{sec:discussion}

McCormack et al. \cite{mccormackHigherOrderSpatialImpulse2020} found that the post-equalized Spatial Decomposition Method (SDM) using pseudo-intensity vectors (PIVs) underperforms compared to spatial impulse response rendering (SIRR) in simulated environments. Improved performance of SDM is observed when PIV-based direction of arrival (DOA) estimation incorporates band-limitation \cite{zaunschirmBRIRSynthesisUsing2018,zaunschirmBinauralRenderingMeasured2020}. More recently, McCormack et al. \cite{mccormackSpatialReconstructionBasedRendering2023} found that SDM, utilizing time difference of arrival (TDOA) based DOA estimation, shows comparable performance to Higher-Order SIRR (HO-SIRR) for stationary stimuli in a small simulated room, but exhibits reduced performance for transient stimuli.

Our study finds similar spatial fidelity for HO-SIRR, and SDM employing TDOA and PIVs for DOA estimation. Although objective measurements showed HO-SIRR to overestimate reverberation time by more than 100 ms, potentially impacting spatial fidelity for non-stationary stimuli, subjective evaluations did not reflect this. The present study stands out by using real-life measurements, contrasting with previous studies' simulated conditions. This offers a more realistic context, directly relevant to practical room acoustics applications.

The performance of SDM with optimizations for binaural rendering (BSDM) depends on the stimulus type. The Bongo stimulus adversely impacted spatial and timbral fidelity, suggesting that BSDM's equalization quality might be compromised by suboptimal reverberation time estimation in low-reverb environments \cite{helmholzPredictionPerceivedRoom2022}. While objective measurements confirm RTmod's efficacy, showing the reverberation time's mean absolute error (MAE) within the just noticeable difference (JND) range, the observed poor performance may be linked to artifacts related to RTmod, appearing in transient sounds \cite{amengualgariOptimizationsSpatialDecomposition2021}. 

BSDM significantly underestimated $\text{1-IACC}_{L3}$, a key metric for listener envelopment \cite{hidakaInterauralCrossCorrelation1995}. This underestimation surpasses the established JND \cite{iso3382-1AcousticsMeasurementRoom2009}, suggesting a potential reduction in listener envelopment relative to the reference.

Amengual et al. observed that BSDM's perceived plausibility was rated similarly to that of real loudspeakers \cite{amengualgariOptimizationsSpatialDecomposition2021}. However, in an ITU-R BS.1116-3-compliant listening room, BSDM shows lower consistency and robustness compared to the standard, unoptimized SDM. This finding, alongside insights from our pilot study---which focused exclusively on DOA enforcement for direct sound and band-limited DOA estimation---suggests that more fundamental optimizations from BSDM could lead to a more robust SDM framework, especially when compared to the full BSDM including RTmod+AP optimization.

Both TDOA and PIVs can be equally effective DOA estimators in SDM when a sufficient number of microphones is utilized for TDOA estimation. In band-limited PIV-based DOA estimation, the condition utilizing an omnidirectional pressure signal demonstrates superior performance compared to the TDOA-based algorithm with a six-microphone array. McCormack et al. \cite{mccormackHigherOrderSpatialImpulse2020} and Ahrens \cite{ahrensPerceptualEvaluationBinaural2019} conducted studies that did not incorporate band limitation in the DOA estimation with the PIV method. In their findings, Ahrens demonstrated that the SDM variant utilizing the PIV method, without band limitation, tended to cause a larger perceived difference in the presented auralizations when compared to most other array geometries, with the dummy head reference serving as the baseline for comparison. Similarly, McCormack et al. observed that the SDM utilizing PIV-based DOA estimation without band limitation, significantly underperformed in comparison to any variant of the SIRR. Zaunschirm et al. \cite{zaunschirmBRIRSynthesisUsing2018} found that applying band limitation to PIV-based DOA estimation significantly enhances SDM performance. Their effective frequency range, from 200 Hz to just below the spatial aliasing frequency of the microphone array, led to SDM results closely mirroring the binaural reference in terms of image width, distance, and diffuseness. These findings align with our study, reinforcing the critical role of band limitation in PIV-based DOA estimation for optimal spatial analysis and synthesis.

We did not expect SDM em32 Omni (E) to match in performance with SDM PIV Omni (I) given the complexity of the time difference of arrival estimation using spherical microphone arrays and the rigid sphere design of em32 \cite{jarrettTheoryApplicationsSpherical2017}. Although the larger number of microphones in em32 suggests a potential for improved TDOA estimation, objective measurements indicate higher errors in ILD, ITD, and reverberation time, aligning em32 more with SDM 6OM1 Omni (E) rather than SDM PIV Omni (I). Previous applications of SDM employed small microphone separations with the smallest inter-mic distance of 17.7 mm \cite{tervoSpatialAnalysisSynthesis2015}. The 26.5 mm inter-mic distance in em32 may suit well for TDOA estimation. Optimizing the TDOA model for microphones on rigid spheres \cite{nikunenTimedifferenceArrivalModel2017} could improve estimation accuracy in em32. 

SDM em32 Omni (E) with 32 sensors outperforms the SDM 6OM1 Omni (F) with six sensors. In contrast, Ahrens' study using different array configurations, including a 12-microphone array, found that a six-microphone array, especially with radii of 50 mm and 100 mm, produced the smallest perceptual differences to the dummy head reference \cite{ahrensPerceptualEvaluationBinaural2019}. The superior performance of the em32 in our study might be attributed to the larger number of microphones in em32 aiding in localization \cite{tervoSpatialDecompositionMethod2013}, as well as a higher spatial aliasing frequency which could positively impact the TDOA estimation \cite{benestyMicrophoneArraySignal2008}.

Using the em32 array with a dedicated pressure signal---obtained from a separate high-quality omnidirectional microphone---seems to give substantial benefits for SDM. Amengual Gari et al.'s \cite{amengualgariSpatialAnalysisAuralization2017} find the benefit of an omnidirectional microphone over an Ambisonic microphone being small, suggesting the influence of equipment and setup on the auralization quality to be more significant.

The spatial and timbral fidelity similarities between HO-SIRR (G) and SDM PIV (H) methods, along with the improvements observed between SDM PIV (H) and SDM PIV Omni (I), suggest that adding a dedicated pressure microphone may improve SIRR methods, as noted in \cite{merimaaSpatialImpulseResponse2005}.

\section{Conclusion} \label{sec:conclusion}
The study evaluated spatial analysis and synthesis methods, specifically SDM, BSDM, and HO-SIRR, along with their variations, in creating synthetic binaural room impulse responses (BRIRs) of an ITU-R BS.1116 compliant listening room. The objective was to achieve auralization perceptually similar to the KU100 dummy head in terms of spatial fidelity and timbral fidelity, using Zieliński's fidelity attribute \cite{zielinskiComparisonBasicAudio2005}. Certain SDM configurations exhibited spatial fidelity levels comparable to those of HO-SIRR, while BSDM suffered from artifacts in the temporal structure of stimuli. A larger number of microphone capsules enhanced the performance of SDM methods, and a separate pressure signal improved timbre. The results are expected to more generally describe the performance of synthetic binaural room impulse responses in small rooms, such as critical listening rooms and audio mixing rooms. Future work can include the study of other rooms, additional subjective dimensions and new systems like 4D-ASDM \cite{hoffbauerFourDirectionalAmbisonicSpatial2022} or REPAIR \cite{mccormackSpatialReconstructionBasedRendering2023}.

\section*{Acknowledgments}
The research presented in this paper was funded by Genelec Oy and the University of Huddersfield. The authors would like to thank everyone who took part in the listening test.

\bibliography{mybib}

\newpage

 




\vfill

\end{document}